# EINSTEIN Y LA VANGUARDIA ARTÍSTICA DE COMIENZOS DEL SIGLO VEINTE: ¿PUNTOS DE CONTACTO?[*]

José X. Martini
Asociación Ciencia Hoy

Cierta tradición de la historia del arte vinculó algunas escuelas pertenecientes a la vanguardia de la producción visual de las primeras décadas del siglo XX con contemporáneas concepciones de la matemática y la física teórica. Una de las asociaciones postuladas con mayor frecuencia es la que habría existido entre cubismo y relatividad, o más precisamente, entre los mundos conceptuales de Picasso y Einstein. También se asoció la relatividad con el futurismo artístico. Los vínculos sugeridos consisten en similitud (en su forma débil) o en identidad (en su forma fuerte) en cuanto al modo de concebir el espacio, el tiempo y la simultaneidad. La cuestión, por otro lado, se inscribe en un marco teórico de cierta complejidad, acerca de las relaciones de ciencia y arte.

Entre los más exitosos difusores de estas teorías se contó Sigfried Giedion (1883-1968), un historiador suizo del diseño y la arquitectura, discípulo de Heinrich Wölfflin, que, en 1938-39, fue profesor visitante en la universidad de Harvard y, en 1941, publicó en inglés un libro con una síntesis de las clases y seminarios que dio en tal carácter. Lo llamó *Espacio, tiempo y arquitectura*, y le agregó el epígrafe *El crecimiento de una nueva tradición*.[1] La obra, atractivamente impresa e ilustrada, no estaba dirigida a especialistas sino al público culto interesado en el arte, y tuvo inmenso éxito: alcanzó cinco reediciones y más de una docena de impresiones, sin contar diversas traducciones, incluso al castellano. Fue por años bibliografía obligada en las escuelas de arquitectura y diseño. Su título, quizá como una primera indicación del pensamiento del autor, es una paráfrasis de *Espacio, tiempo y gravitación*, un libro de divulgación de la teoría general de la relatividad escrito por el astrónomo británico Arthur Eddington. La parte VI de la obra de Giedion se llamó 'Espacio-tiempo en el arte, la arquitectura y la construcción', y el primer capítulo de esta, 'La nueva concepción del espacio: espacio-tiempo'. Los argumentos del autor eran:

> *El espacio tridimensional del Renacimiento es el de la geometría euclidiana. Pero alrededor de 1830 se creo una nueva clase de geometría, que difería de la de Euclides por emplear más de tres dimensiones. [...] Como el científico, el artista ha llegado a reconocer que las concepciones clásicas del espacio y los volúmenes son limitadas y unilaterales. [...] Las escaleras de los niveles superiores de la torre Eiffel se cuentan entre las expresiones*

---

*arquitectónicas más tempranas de la continua interpenetración del espacio interior y exterior. [...] Y en el arte moderno, por primera vez desde el Renacimiento, una nueva concepción del espacio condujo a una ampliación consciente de nuestra manera de concebir el espacio. Fue en el cubismo donde esto se percibió más plenamente.*

En otros pasajes del mismo libro afirmó:

*El cubismo rompe con la perspectiva renacentista. Mira los objetos relativamente, es decir, desde varios puntos de vista, ninguno de los cuales tiene autoridad exclusiva. Y al disecar los objetos de esa manera, los ve simultáneamente de todos lados: de arriba y abajo, de adentro y afuera. Rodea y penetra los objetos. Así, a las tres dimensiones del Renacimiento que se mantuvieron vigentes por tantos siglos, se agrega una cuarta, el tiempo. El poeta Guillaume Apollinaire fue el primero en reconocer y expresar ese cambio, alrededor de 1911. El mismo año vio la primera exposición cubista en el Salón de los independientes. Si se considera la historia de los principios con los que rompieron, bien se comprende que los cuadros hubiesen sido considerados una amenaza a la paz pública y fueran objeto de comentarios en la cámara de Diputados.*

*La presentación de objetos desde varios puntos de vista introduce un principio íntimamente ligado con la vida moderna, la simultaneidad. Es una coincidencia temporal que Einstein haya comenzado su famoso trabajo de 1905, Electrodinámica de los cuerpos en movimiento, con una cuidadosa definición de la simultaneidad.*

En estos párrafos Giedion dejó vinculadas las nociones de geometrías no euclidianas y de espacios de más de tres dimensiones con las construcciones de acero y con el cubismo. Lo hizo por una asociación libre que apela a la imaginación más que por un razonamiento riguroso que llega a conclusiones fundadas. Después de todo, las escaleras metálicas en caracol ubicadas en lo alto de la torre Eiffel no son conceptualmente muy distintas (ni son, por cierto, estéticamente más creativas) que muchas escaleras medievales o barrocas de piedra. Justificó la mención del cubismo recurriendo al concepto de espacio-tiempo, e inmediatamente evocó a Einstein y la relatividad especial, aunque, nuevamente, en términos de libre asociación de ideas y no de argumentos que desembocan en conclusiones. Después declaró:

*En la primera década del siglo, las ciencias físicas fueron profundamente sacudidas por un cambio interno, el más revolucionario quizá desde Aristóteles y los pitagóricos. Dicho cambio se refirió, sobre todo, a la noción de tiempo. [...] Vino un modo distinto y nuevo de considerar el tiempo, con implicancias del mayor significado, cuyas consecuencias no se puede hoy minimizar o ignorar. [...] Hermann Minkowski, el gran matemático, hablando ante la Sociedad de Naturalistas, proclamó por primera vez con toda certeza y exactitud este cambio fundamental de concepción. [...] Concurrentemente, las artes se ocupaban del mismo problema. [...] El cubismo y el futurismo intentaron ampliar nuestra visión óptica mediante la introducción de la nueva unidad de espacio-tiempo en el lenguaje del arte.*

Minkowski, que había sido profesor de Einstein en Zürich, propuso en 1908 conceptos matemáticos innovadores adaptados a las nociones einsteinianas de la relatividad especial publicada tres años antes. Explicó que esos conceptos de espacio y tiempo *surgieron del suelo* *de la física experimental*, *en lo que reside su fuerza* (aunque, posiblemente, la relación más directa haya sido con la física teórica, no la experimental).

Estas relaciones entre física y arte de vanguardia no fueron señaladas por un físico o por un artista sino por un crítico e historiador, que las expuso con brillo e, incluso, espectacularidad, si bien con escaso fundamento documental. Giedion dio a conocer su análisis, por otra parte, unas tres décadas después de haberse producido los hechos a los que se refirió. Cuando estos acontecían, en los ámbitos en que se movían los artistas no habían llegado noticias de las nuevas teorías físicas, ni de las repercusiones de estas en la matemática.

Einstein era una figura relativamente desconocida, aun para sus pares, ya que en 1908 ni siquiera era profesor universitario. Se incorporó al sistema académico formal solo en 1909, en Zürich, de donde pasó a Praga en 1911, retornó a Zürich en 1912, y llegó al centro mundial de la física de entonces, a Berlín, de la mano de Max Planck, en 1914. Se puede decir que para ese momento era conocido por la vanguardia de su disciplina y por los círculos ilustrados de Berlín, pero su nombre no alcanzó mayormente al público general y, por ende, a los cenáculos artísticos de avanzada de París, donde tuvo lugar la revolución cubista, hasta fines de esa década, cuando, aprovechando un eclipse de sol que aconteció en 1919, el mencionado Eddington dirigió una expedición al África costeada por la *Royal Society* de Londres y comprobó experimentalmente las predicciones de la relatividad general acerca de la desviación de la luz por efecto de la gravedad. Esto fue divulgado por los medios sobre la base de conferencias, declaraciones y escritos de los propios físicos, como el indicado libro de Eddington, y causó sensación. En 1921, el renombre de Einstein se incrementó con la obtención del premio Nobel (por una contribución ajena a la relatividad).

Solo en abril de 1921, una publicación propia de los medios artísticos, *L'Esprit Nouveau*, sacó un artículo de Paul Le Becq titulado 'À propos des théories d'Einstein' (A propósito de las teorías de Einstein), lo que indica que para entonces las noticias de estas teorías habían llegado al conocimiento de dichos medios. Se puede pensar, pues, que el origen, sugerido por Giedion y por algunos otros, de las nuevas ideas artísticas encarnadas en el cubismo y el futurismo, si fue ajeno al arte mismo, no estuvo en la física de Einstein sino en otra parte.

No debe darse por sentado, sin embargo, que las ideas artísticas del cubismo y el futurismo hayan venido de afuera del arte mismo. Si así hubiese sido, se hubiera tratado de una situación anormal en la historia del arte: una excepción y no la regla, y sería necesario demostrar que efectivamente tal cosa sucedió. No basta con afirmarlo: la carga de la prueba debe recaer en quien postule la anomalía. Timothy J Clark, un crítico marxista de arte, afirmó que *la pintura raramente come bien con las sobras de la ciencia*. [En el cubismo] *se alimenta principalmente de sí misma*.[2]

Giedion no fue el único a señalar una relación entre cubismo y relatividad. Antes que él, Thomas J. Craven se había referido a cierto paralelismo entre ambos, y no mucho después de Giedion lo hizo Walter Isaacs. También hubo desacuerdos, como el de Joachim Weyl, quien sostuvo que el cubismo se relacionaba mejor con la ciencia de Descartes y Newton. Edward Fry indicó en un libro clásico que, *si bien se podría trazar legítimamente un paralelismo metafórico grueso* [entre cubismo y relatividad], *parecería peligroso llevarlo demasiado lejos*. También usó la expresión 'metáfora' Robert Motherwell, el editor de una traducción al inglés de Guillaume Apollinaire, acerca de quién un genetista escocés llamado Conrad H. Waddington sostuvo que mostraba *entusiasmo por una ciencia entendida a medias*, y que sus opiniones eran *tonterías* (nonsense) *típicas del absoluto disparate* (gibberish) *que producen los críticos de arte*.[3]

Por otro lado, el concepto de la cuarta dimensión en el que Giedion centró su argumentación había cobrado vida propia desde hacía bastante, en la literatura no científica y en la imaginación del público culto. Más de una década antes de que tomara forma el cubismo, y diez años antes de que Einstein diera a conocer la teoría especial de la relatividad, H.G. Wells publicó su *Máquina del tiempo*, una novela que contenía la siguiente afirmación: *Hay realmente cuatro dimensiones*, *tres de las cuales llamamos los tres planos del espacio y la cuarta*, *el tiempo*. Se podría asimismo citar el antecedente un poco más remoto de la novela *Flatland*, publicada en Londres en 1884, una sátira social sobre un mundo de dos dimensiones escrita por Edwin A. Abbott, un director de escuela enteramente ajeno al mundo de las matemáticas. El fantasma de Canterville de Oscar Wilde (1887) utilizó *la cuarta dimensión del espacio para huir*, mientras que Rudyard Kipling escribió un cuento llamado *Un error en la cuarta dimensión* (1894). Marcel Proust habló de *un espacio de cuatro dimensiones* en *Du côté de chez Swann* (1913). El concepto de cuarta dimensión también tuvo su vertiente mística y ocultista, que se extendió a la teosofía.

Un concurso con un premio de 500 dólares ofrecido en 1909 por la revista de divulgación *Scientific American* a la mejor explicación popular de la cuarta dimensión atrajo más de 200 ensayos y dio lugar a la publicación de un volumen con una veintena de ellos (ninguno menciona a Einstein o la relatividad).[4] John Richardson sostuvo que *para el fin del siglo* XIX, *tanto la ficción seria como la literatura pseudo científica habían preparado a cierto segmento del público lector a que aceptara la idea del tiempo como una dimensión. La aparición del mundo del espacio-tiempo de la física moderna confirmó en la mente de esa gente la exactitud de la noción*. El libro de Richardson es uno de los dos de relativamente reciente publicación que contienen un análisis sistemático del tema que trata este escrito; el otro, que quizá pueda considerarse la obra definitiva de erudición sobre el asunto, fue escrito por Linda Henderson.[5]

Richardson recordó que un cubista francés de segunda línea, Jean Metzinger, opinó en 1912 que el cubismo *representaba el tiempo igual que lo hacían las nuevas teorías*, *como una dimensión*, y mostraba acontecimientos sucesivos como si fuesen simultáneos. Pero tampoco parece que esta mención de 'nuevas teorías' aluda a la física, sino a la noción de un espacio de cuatro dimensiones, es decir a la geometría, lo que coincide con lo afirmado en el libro de Apollinaire de 1913: *Los nuevos pintores no se propusieron ser geómetras, como no lo hicieron sus antecesores. Pero se puede decir que la geometría es a las artes plásticas lo que la gramática es al arte del escritor. Hoy los científicos no se ajustan más a las tres dimensiones de la geometría euclidiana. Los pintores se vieron conducidos muy naturalmente y por intuición*, *por así decirlo*, *a preocuparse por nuevas posibles medidas del espacio que*, *en el lenguaje de los talleres modernos*, *se designaban en conjunto y brevemente por el término 'cuarta dimensión'*. La evidencia sugiere que tales ideas pudieron llegar a los pintores, entre otras vías, por Maurice Princet, un actuario de seguros integrante del entorno de Picasso, quien habría estado al tanto de una obra de divulgación publicada en París en 1902 por Poincaré, en la que este había escrito: *No hay espacio absoluto* [...] *No hay tiempo* *absoluto*.[6] Richardson agregó que la opinión de Metzinger *se puso desde entonces crecientemente de moda y se convirtió, en realidad, en una muletilla de la crítica contemporánea*. Edward Fry, por su lado, sostuvo que las menciones de geometrías no-euclidianas y la cuarta dimensión por Apollinaire *solo sirvieron para oscurecer la comprensión del cubismo mediante un misticismo pseudocientífico*.

Por lo que se está viendo, sin embargo, tal muletilla, basada en nociones construidas por la literatura y la propia crítica, no parece directamente fundamentada en las teorías físicas, ni se deduce de las manifestaciones de los principales artistas. La representación simultánea y contigua de acontecimientos u objetos que, en la realidad, no se hubiesen podido ver uno al lado del otro, por estar separados en el tiempo o el espacio, fue recurso usual en el arte a partir del Renacimiento, lo que la acercaba más al contexto cultural de la física clásica que al de la moderna (como lo sostuvo en 1943 el citado Weyl), mientras que la relatividad especial de Einstein había, precisamente, puesto en cuestión la simultaneidad absoluta de acontecimientos espacialmente separados.

Picasso, por su lado, afirmó en 1923 que *matemática, trigonometría, química, psicoanálisis, música y qué se yo qué fueron relacionadas con el cubismo, para darle una fácil interpretación. Todo fue pura literatura, por no decir tontería, que ocasionó malos resultados y cegó a la gente con teorías*.[7]

Uno de los documentos más interesantes sobre las supuestas relaciones entre relatividad y arte moderno de vanguardia fue hecho público por Paul Laporte, un profesor universitario de historia del arte que se ocupó del tema en 1945 y produjo un ensayo que tituló 'Cubismo y la teoría de la relatividad'. Antes de publicarlo sometió el texto al propio Einstein, que lo leyó con la detención necesaria como para escribir en mayo de 1946 una cuidadosa respuesta, en alemán, que, traducida al inglés por el destinatario, alcanzó las 500 palabras.[8]

Laporte había sostenido que existían *ciertas analogías entre esas manifestaciones por otra parte altamente divergentes de la cultura contemporánea* (el cubismo y la teoría de la relatividad). Su argumento había sido que ambas abandonaron la vieja modalidad de prestar atención a los objetos sin considerar el modo de observación y, en cambio, *tuvieron en cuenta relaciones y la simultaneidad de varios puntos de vista*. Concluyó que, como consecuencia, se produjeron, respectivamente, *una aparente distorsión o disolución de los cuerpos en la pintura y la famosa convertibilidad de masa y energía en la teoría de la relatividad*. Y agregó: *Espacio y tiempo constituyeron el continuo espacio-tiempo, que, por su parte, no era sino una forma de experiencia humana. En la pintura, el símbolo del espacio-tiempo fue el dibujo geométrico característico del cubismo, solo experimentado por el observador del cuadro*.

Einstein comenzó su respuesta afirmando: *Encuentro su comparación poco satisfactoria*. Luego señaló que dicha comparación evidenciaba *una incorrecta comprensión de la teoría de la relatividad, y que ello podía deberse a los intentos de popularizarla*. Hacia el final del escrito mencionó a Picasso y afirmó (quizá como forma educada de tomar distancia de su pintura) que la posibilidad de atribuir a su obra carácter artístico dependería de *la historia artística previa del observador*. Y como última frase puso, lapidariamente, *este nuevo 'lenguaje' artístico no tiene nada en común con la teoría de la relatividad*.

Es posible que la discreta manifestación hecha por Einstein de desinterés por la obra de Picasso proporcione una clave importante para interpretar su carta a Laporte —y, en última instancia, su descalificación de los argumentos de este—, sobre todo si se la relaciona con las densas disquisiciones que componen el grueso de la carta, referidas a las respectivas naturalezas de ciencia y arte. Tales disquisiciones indican, a nuestro juicio, no solo un genuino esfuerzo del físico por entender la argumentación de Laporte, sino, también, de reflexionar sobre el arte e, incluso, de apreciarlo, si bien uno casi se atreve a afirmar que esta última posibilidad seguramente se encontraba entorpecida por el fuerte predominio de sus hábitos e inclinaciones —y, obviamente, su extraordinaria capacidad— para la construcción de explicaciones racionales rigurosas.

Einstein postuló la existencia de una similitud entre ciencia arte, que, según él, residiría en el esfuerzo de ambas por crear una unidad clara y distinta partiendo de algo caótico (*unübersichtlich*). Aclaró que el principio ordenador es de orden *lógico* en la ciencia, y de tipo *inconsciente y tradicional* en el arte, en cuyas formas sencillas, basadas en una regularidad evidente, es claro de percibir y resulta tan forzoso como la conclusión de un razonamiento matemático, pero en formas artísticas más complejas resulta de percepción menos obvia. Uno casi adivina entre líneas el desconcierto del físico ante las manifestaciones artísticas de la vanguardia, o de lo que por entonces solía también llamarse el arte moderno. Lo mismo hace pensar la afirmación de que una obra de arte solo puede ser apreciada por aquellos en quienes está vivo el correspondiente principio ordenador.

¿Resulta la descalificación de Einstein suficiente para concluir que las relaciones establecidas por Giedion y Laporte carecen de fundamento? Y las palabras de Picasso que se citaron, ¿no refuerzan esa descalificación? Depende del alcance que se pretenda dar a los vínculos entre las nuevas teorías físicas y el arte moderno. Sin duda, por falta de sustento documental sería muy difícil defender la existencia de alguna forma de causalidad estricta en una u otra dirección, algo que a veces parecen insinuar determinados análisis algo ligeros. Pero los autores más serios, como los citados y otros, no sugieren que haya existido tal cosa. Por lo general no indican mucho más que posibles influencias de ciertas ideas científicas sobre determinadas escuelas artísticas (nunca la recíproca), o incluso conexiones más débiles. Sin embargo, como lo señaló explícitamente Einstein, esas ideas no suelen ser las específicas de una disciplina académica formal sino de sus formas vulgarizadas o, incluso, las imágenes literarias que esas formas pudieron provocar. Por otro lado, el vínculo parece haberse establecido más con conceptos de las geometrías no euclidianas que con las teorías de la física.

La mayoría de los comentaristas sugiere paralelismos relativamente laxos entre arte y ciencia, con escaso valor explicativo, salvo que —considerando el conjunto de la cultura de una época— se mirara a arte y ciencia como dos de las varias expresiones de un más bien inasible estilo o espíritu de los tiempos. La noción de tal espíritu o *Zeitgeist* —un concepto cuyas raíces decimonónicas están en el romanticismo alemán y en los filósofos idealistas de ese origen— puede dar lugar a una rica figura literaria, pero no proporciona una categoría analítica que permita avanzar demasiado en la comprensión de las supuestas relaciones. Es más, antes que ayudar a entender los cambios en la ciencia o las artes, analogías como las señaladas han tendido a menudo a crear una cortina de mistificación que oculta sus alcances. Kenneth Clark estableció otro paralelismo laxo cuando indicó que, en su opinión, *la* *incomprensibilidad de nuestro nuevo cosmos es*, *en última instancia*, *la razón del caos en el arte moderno*, cuyo corolario podría ser que la comprensión proporcionada por la física relativista explicaría la forma que tomó ese arte.[9]

En cuanto a la conexión específica entre arte de vanguardia de inicios del siglo XX y las nuevas teorías de la física de entonces, en palabras de Richardson, que de alguna manera puso las cosas en su lugar, las *analogías son tan engañosas como ubicuas. Tergiversan tanto el cubismo como la física moderna. Dado que surgen con alguna frecuencia en el marco de la historia de la cultura y la crítica de arte*, *y que se presentan como el principio unitario de la modernidad*, *es importante dedicar alguna energía a desacreditarlas*. Unas páginas más adelante en el libro que hemos citado, el mismo autor agregó: *Curiosamente*, *hubo desde el principio una tendencia a aplicar las ideas de la física al cubismo sin tratar de verificarlas ni de comprobar sus aplicaciones. Y a medida que se consuma una flagrante falsedad*, *ella comienza a sonar cada vez más razonable*, *dado que se la oye con creciente frecuencia*. Como conclusión, escribió: *...se puede sostener que toda la noción de una conexión hermética entre la teoría de Einstein y el cubismo es falsa. De hecho*, *el mismo Einstein la anuló.* [...] *El cubismo no tiene nada que ver con la teoría de la relatividad. Este es el final de la cuestión*. Dicho eso dejó, de todos modos, una puerta abierta al espíritu de los tiempos, pues manifestó: *Tal afirmación*, *sin embargo*, *no niega que pudo haber existido una relación significativa entre estilo pictórico y ciencia moderna*, *o entre cubismo y la cultura general que recibió tan vasta influencia de la ciencia*. La frase no es de las más afortunadas, porque insinúa tanto la afirmación categórica (*una relación significativa*) como la duda (*pudo haber existido*), y tiene un grado tal de generalidad que poco ilumina. Quizá habría que considerarla, en las palabras de Picasso, *pura literatura*.

En síntesis, si a la luz de lo dicho y en línea con el título de este ensayo, retornamos a los posibles vínculos entre Einstein y la vanguardia artística de principios del siglo XX, parece inevitable concluir que ellos fueron, sencillamente, inexistentes. Por un lado, las artes visuales (a diferencia de la música, hasta donde sabemos) no eran objeto de interés para el físico, quien, además, no parecía predispuesto ni suficientemente versado para apreciar sus manifestaciones avanzadas. Por otro lado, las teorías de Einstein estaban fuera de la órbita de los artistas, para no hablar de sus posibilidades de comprensión, aunque haya indicios firmes de que ciertas repercusiones de ellas en los medios periodísticos y literarios pueden haber estimulado la imaginación de determinados creadores.

Se conoce una circunstancia en la que Einstein estuvo directamente expuesto a la vanguardia artística, aunque no se trató de artes visuales sino de arquitectura. Sucedió en Berlín hacia 1920, cuando ya se había producido la mencionada comprobación experimental de una de las predicciones de la relatividad general y, como consecuencia, el físico ya era famoso. Además del desvío de la luz por la fuerza de gravedad, que fue el efecto verificado por Eddington, la relatividad también predecía un corrimiento hacia el rojo del espectro de la luz solar, efecto que se propuso medir en Berlín el astrónomo Erwin Finlay Freundlich (1885-1964). Para hacerlo, se requería montar en el Instituto Astrofísico (*Astrophysikalisches Institut*) de Potsdam un telescopio especial acoplado a un espectrógrafo, lo que llevó a la necesidad de erigir un edificio que los albergara. Así se construyó el observatorio solar que terminó llamándose la torre Einstein (*Der Einsteinturm*), una de las obras emblemáticas de la temprana arquitectura moderna, debida al arquitecto Erich Mendelsohn.[10]

Nacido en 1887 en Prusia oriental, Mendelsohn había llegado a Berlín en 1915, tres años después de terminar sus estudios de arquitectura en Munich, donde había frecuentado a los integrantes del grupo artístico de vanguardia *Der blaue Reiter* (El jinete azul). Conoció a Freundlich en los circuitos musicales de la capital imperial (que sería capital de la república de Weimar con la creación de esta después de la guerra), y se enteró así de la iniciativa de construir el observatorio, para el cual propuso algunas ideas de diseño antes de recibir, en 1920, el encargo formal de proyectarlo. El edificio, que se inauguró en diciembre de 1924, fue llamado *la construcción más significativa del expresionismo alemán*, mientras que para un crítico, *más que un laboratorio*, [Mendelsohn] *creó un monumento*. En un artículo publicado en 1969, Freundlich lo describió como *un monumento a la memoria del trascendente alcance de la teoría de la relatividad para el desarrollo de la física*.

Como con el cubismo, también en este caso hubo quien buscó los vínculos entre las inusuales características del edificio y los revolucionarios postulados de la relatividad. Un asistente de Freundlich que con el tiempo dirigió el observatorio, Harald von Klüber (1901-1978), solía sugerir a visitantes que recorrían los diversos pabellones del Instituto Astrofísico, que aquellos construidos en el siglo XIX reflejaban los conceptos de la geometría euclidiana y las nociones clásicas de la estructura de la materia, mientras la torre de Mendelsohn evocaba la visión de Einstein del espacio y la equivalencia de materia y energía, si bien, en palabras de un crítico actual, *tenía conciencia de que ello carecía estrictamente de sentido y que la creación artística de un arquitecto no necesita una explicación racional*.[11] La posición de Einstein de descalificar esa clase de especulaciones quedó sucintamente expresada en una frase coloquial que le respondió por carta a Mendelsohn cuando este le hizo llegar los pasajes de Giedion que citamos a comienzos de este escrito: *Es sencillamente una pedantería sin base racional alguna*.[12]

---

[1] Siegfried Giedion, *Space, Time and Architecture. The growth of a new tradition*, Harvard University Press, 1941.

[2] Timothy J Clark, *Farewell to an Idea: Episodes from the History of Modernism*, Yale University Press, 1999.

[3] Las referencias de las citas de este párrafo son: Thomas Jewell Craven, 'Art and Relativity', *The Dial*, 5, 70, 1921; Walter Isaacs, 'Time and the fourth dimension in painting', *College Art Journal*, 2, 1, noviembre 1942; Joachim Weyl, *College Art Journal*, 2, 2, enero 1943; Edward Fry, *Cubism*, McGraw-Hill, Nueva York 1966; Guillaume Apollinaire, *The Cubist Painters*, Nueva York, 1949, traducción de *Les peintres cubistes. Méditations esthétiques*, Eugène Figuière, Paris 1913, y Conrad H Waddington, *Behind Appearance: A Study of the Relations between Painting and the Natural Sciences in this Century*, University of Edinburgh Press, 1969.

[4] Henry P. Manning, *The Fourth Dimension Simply Explained. A Collection of Essays Selected From Those Submitted in the Scientific American's Prize Competition*, Munn & Co, Nueva York 1910. El texto completo podía consultarse (agosto 2005) en *http://etext.lib.virginia.edu/toc/modeng/public/ManFour.html*. El ensayo ganador, escrito por Graham Denby Fitch, había salido en la revista el 3 de julio de 1909.

[5] John Adkins Richardson, *Modern Art and Scientific Thought*, University of Illinois Press, 1971. Linda Dalrymple Henderson, *The Fourth Dimension and Non-Euclidean Geometry in Modern Art*, Princeton University Press, 1983.

[6] Jules-Henri Poincaré, *La Science et l'hypothèse*, Flammarion, París, 1902. El texto completo podía consultarse (agosto 2005) en: *http://www.univ-nancy2.fr/poincare/bhp/sh.html*.

[7] Herschel B Chipp, *Theories of Modern Art: A Source Book by Artists and Critics*, University of California Press, Berkeley 1968, p.265.

[8] Laporte publicó su artículo original en dos partes ('The Space-Time Concept in the Work of Picasso', *Magazine of Art*, 41, 1:26, enero 1948, y 'Cubism and Science', *The Journal of Aesthetics and Criticism*, 7, 3:243, marzo 1949). En 1966 dio a conocer un tercer artículo, en el que transcribió su traducción inglesa de la respuesta de Einstein ('Cubism and Relativity', con una carta de Albert Einstein del 4 de mayo de 1946, *Art Journal*, 25, 3:246-248, primavera de 1966).

[9] Kenneth Clark, *Civilization*, Harper & Row, Nueva York, 1969.

[10] JX Martini, '*Der Einsteinturm*. Física, astronomía, arquitectura y la financiación de la ciencia en la Alemania de entreguerras', *Ciencia Hoy*, 41:54-64, 1997.

[11] Barbara Eggers, 'Der Einsteinturm–die Geschichte eines Monumentes der Wisseschaft', en AAVV, *Der Einsteinturm in Potsdam. Architektur und Astrophysik*, Ars Nicolai, Berlín 1995.

[12] *Es ist einfach Klugscheißerei ohne jede vernünftige Basis*. Carta de Einstein a Mendelsohn, 13 de noviembre de 1941, Kunstbibliothek, Berlín, citada por **Joachim Krausse**, '**Von Einsteinturm zum Zeiss-Planetarium. Wissenschaftliches Weltbild und Architectur**', en AAVV, *op. cit*. 1995, p.106.